# Infrared Nanoimaging of Hydrogenated Perovskite Nickelate Synaptic Devices

Sampath Gamage[1], Sukriti Manna[2,3], Marc Zajac[4], Steven Hancock[5], Qi Wang[6], Sarabpreet Singh[1], Mahdi Ghafariasl[1], Kun Yao[7], Tom Tiwald[8], Tae Joon Park[6], David P. Landau[5], Haidan Wen[4,9], Subramanian Sankaranarayanan[2,3], Pierre Darancet[2,10], Shriram Ramanathan[6,11] and Yohannes Abate[1*]

[1]Department of Physics and Astronomy, University of Georgia, Athens GA 30602, USA

[2]Center for Nanoscale Materials, Argonne National Laboratory, Lemont, IL 60439, USA

[3]Department of Mechanical and Industrial Engineering, University of Illinois, Chicago, IL, 60607, USA

[4]Advanced Photon Source, Argonne National Laboratory, Lemont, Illinois 60439, USA

[5]Center for Simulational Physics and Department of Physics and Astronomy, University of Georgia, Athens GA 30602, USA

[6]School of Materials Engineering, Purdue University, West Lafayette, IN 47907, USA

[7]School of Electrical and Computer Engineering, University of Georgia, Athens GA 30602, USA

[8]J.A. Woollam Co., Inc., Lincoln, NE 68508, USA

[9]Materials Science Division, Argonne National Laboratory, Lemont, Illinois 60439, USA

[10]Northwestern Argonne Institute of Science and Engineering, Evanston, IL 60208, USA

[11]Department of Electrical & Computer Engineering Rutgers, The State University of New Jersey, Piscataway, NJ 08854, USA

*Corresponding author E-mail: yohannes.abate@uga.edu

## Abstract

Solid-state devices made from correlated oxides such as perovskite nickelates are promising for neuromorphic computing by mimicking biological synaptic function. However, comprehending dopant action at the nanoscale poses a formidable challenge to understanding the elementary mechanisms involved. Here, we perform operando infrared nanoimaging of hydrogen-doped correlated perovskite, neodymium nickel oxide (H-NdNiO$_3$) devices and reveal how an applied field perturbs dopant distribution at the nanoscale. This perturbation leads to stripe phases of varying conductivity perpendicular to the applied field, which define the macroscale electrical characteristics of the devices. Hyperspectral nano-FTIR imaging in conjunction with density functional theory calculations unveil a real-space map of multiple vibrational states of H-NNO associated with OH stretching modes and their dependence on the dopant concentration. Moreover, the localization of excess charges induces an out-of-plane lattice expansion in NNO which was confirmed by in-situ - x-ray diffraction and creates a strain that acts as a barrier against further diffusion. Our results and the techniques presented here hold great potential to the rapidly growing field of memristors and neuromorphic devices wherein nanoscale ion motion is fundamentally responsible for function.

**Main**

Correlated oxides, specifically rare-earth nickelates ($RNiO_3$ where R = rare-earth element), provide a promising platform to configure quantum phenomena at the atomic scale for neuromorphic devices and applications[1-8]. In particular, the metal-insulator transition (MIT) can be modulated by interstitial doping to enable a reconfigurable synaptic unit[8]. However, a detailed understanding of the mechanism of individual synaptic action and control of dopant related phenomena at the nanoscale remains a colossal challenge. This is because changes in charge concentration gradients occur within tiny length scales and minute changes can produce immense effects in the transmitted signal. This presents outstanding challenges in both resolution and sensitivity to localized charge and electronic structure.

In this work, we perform operando infrared nanoimaging of hydrogen-doped neodymium nickel oxide (H-NNO) devices to uncover how an applied electric field (E-field) can disrupt distribution, leading to localized nanoscale phases that govern the device's overall electrical response. We map in real-space at the nanometer length scale the electric-field-driven dopant migration and electronic character of H-NNO thin-film. The hydrogen acts as a donor dopant by donating an electron to the Ni-O orbital manifold and the proton resides as an interstitial. The electron doping results in a massive metal to insulator transition. By controlling the concentration and distribution of the dopant, it is possible to obtain a multitude of resistance states for synaptic function[9-11]. While multi-modal characterization of H-doped nickelate films has been reported[12-14], the nanoscale perturbation of doped regions in an electric-field driven device remains an outstanding problem. Here, we found ordered steady-states that exhibit alternating conducting and insulating stripe phases perpendicular to the field direction due to field-driven proton migration. Using density functional theory (DFT) calculations in conjunction with high-resolution x-ray diffraction (HRXRD) strain mapping and infrared spectroscopy of the vibrational properties, we reveal that this macroscopic state emerges because of the competition between the strain created by the excess electron localization and the field-induced proton drift. By combining ellipsometry dielectric data and empirical calculations we quantify the nanoscale modifications to the dielectric environment due to doping. We found that the migration of protons, driven by an applied field, caused a significant macroscopic increase in the material's size, perpendicular to the direction of migration of the protons, by ~6% due to the buildup of excess charge locally. At the same time, the E-field driven migration also formed insulating barriers that run perpendicular to the flow of current.

**Results**

Figure 1a shows schematic of a biological synapse and its gap junction which enables the exchange of currents[15]. Figure 1b shows schematics of the crystal structure of hydrogen doped $NdNiO_3$ (H-NNO). Pristine bulk NNO is a d-band electron-correlated pseudo-cubic perovskite structure with lattice constant a = 0.3807 nm. The R cations ($Nd^{3+}$) are positioned in the cavities of ordered $NiO_6$ octahedral networks. Figure 1c shows schematics of a gated H-NNO sample, that mimics a biological synapse (Fig. 1a), and the scattering type Scanning Near-field Optical Microscopy (s-SNOM) setup used for nanoimaging and nano-FTIR (Fourier Transform Infrared) spectroscopy. The s-SNOM setup is based on a tapping mode atomic force microscope (AFM) where a metal coated cantilevered probe tip with an apex radius of ~ 20 nm oscillates at a frequency of $\Omega$~ 280 kHz with a tapping amplitude of ~100 nm (see Methods for details).

Here, we demonstrate active manipulation of dopants using E-fields applied between electrodes in contact with the sample. Figure 1d shows a time series of topography and near-field optical amplitude images as a function of applied external field. Since the NNO film is initially partially hydrogen doped, it exhibits a coexistence of the insulating and metallic phases at zero bias as shown in Fig. 1d(v). The near-field amplitude images represent the value of the real part of the dielectric constant; hence the regions with metallic phase (M) give higher signal (yellow) whereas the regions with insulating phase (I) give lower signal (red). Note that the metallic phase (M) corresponds to low hydrogen doping while the insulating phase (I) refers to high level of doping throughout this manuscript. The local phase modulation of H-NNO is the result of the inhomogeneous distribution of the dopant hydrogen atoms donating electrons to the Ni orbitals, modifying the electronic state of the d orbitals in NNO and generating a M/I boundary as reflected in the nanoscale s-SNOM amplitude contrast images (See also Fig. S2).

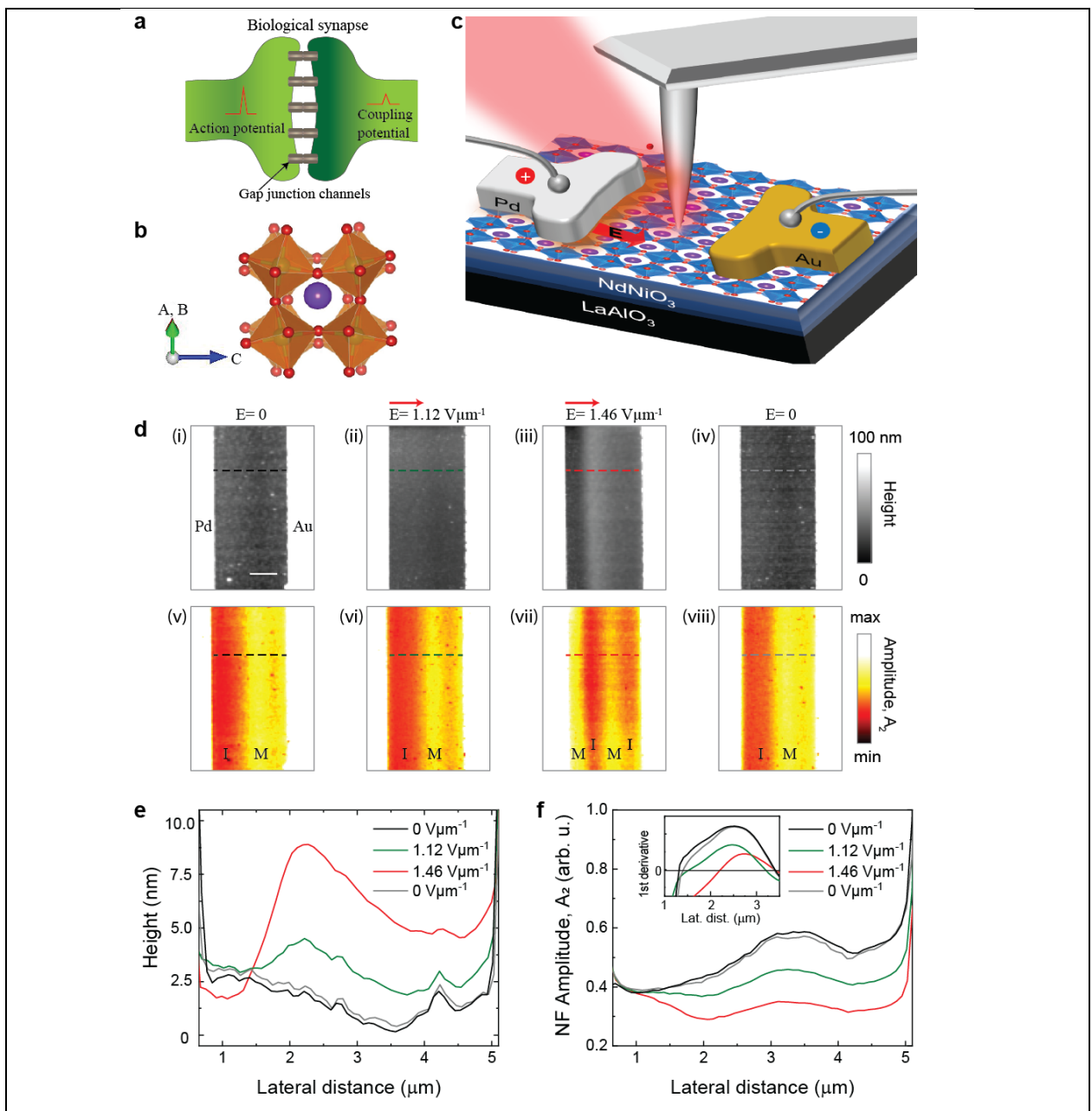

**Fig. 1** Schematics of the experimental setup, nanoimaging and current-electric field relationships. (a) Schematic of a biological synapse that involves the transmission of electrical signals between two neighboring neurons. (b) Schematic of crystal structure of H-doped NNO. Nd atom is in purple at the center, O atoms are at the corners of octahedra and Ni atoms are at the center of each octahedra. (c) Schematics of the s-SNOM experimental setup and H-NNO/LAO synaptic device with Pd and Au electrodes. The metal coated probe tip with apex radius of ~ 20 nm is illuminated by a mid-IR monochromatic laser for single frequency imaging or a broadband light source for nano-spectroscopy. Backscattered light from the tip-sample interface was detected and demodulated at higher harmonics of tip resonance frequency using

interferometry methods as a function of bias voltage-field. (d) Real-space nanoscale imaging of hydrogen migration in topography (i-iv) and s-SNOM amplitude (v-viii) images of doped H-NNO as a function of applied E-field. Topography (b(i-iv)) and applied E-field dependent near-field amplitude images (b(v-viii)) for E= 0, 1.12, 1.46 and 0 Vμm$^{-1}$, respectively. Scale bar in d(i) is 1 μm. (e) Line profiles of topography and (f) amplitude along the lines marked on the topography and amplitude images shown in (d). The inset in (f) shows first derivatives of the amplitude line profiles.

During the field-dependent imaging experiments using an H-NNO device with Pd-Pd electrodes, the polarity and the magnitude of the E-field were kept constant and the recorded images are shown in Fig. 1d. The action of the applied field between the Pd electrodes pointing from left to right, creates two simultaneously measured changes in the sample: *(i)* the sample starts to physically expand as measured by AFM topography and, *(ii)* the dopants are pushed to form a new distribution between the electrodes as revealed by s-SNOM amplitude images. The topography images in Fig. 1d and corresponding line profiles extracted from topography images (see Fig. 1e and SI Fig. S1) reveal a dramatic structural expansion. We quantified this topographic change by measuring the height of the sample before and after E-field application and found up to a ~6% expansion (see more details of topographic height measurement in Fig. S1). When the E-field is turned off, the topography change is reversed and reverts to its original height demonstrating a piezo-like effect. Simultaneously, new M/I phases are created and captured by the s-SNOM amplitude images shown in Fig. 1d. As the strength of the applied E-field is increased, alternating M and I phases appear clearly in the amplitude image (Fig. 1d(vii)). The M or I phases form perpendicular to the current flow axis presenting a transverse barrier to the movement of charge carriers that are driven by the electrical stimulus. This type of resistive switching is quite different from the more common conducting filament generation parallel to the current flow reported for other oxides exposed to E-fields[16-18]. The amplitude image recovers its original M-I contrast when the applied field is turned off, as can be seen by comparing Fig. 1d(v) and Fig. 1d(iii).

The topographic height and the near field contrast changes at each pixel as a function of the applied field are also clearly captured by the line profile curves shown in Fig. 1e and Fig. 1f respectively. Comparison of these curves show a one-to-one correlation of the change in topography to the amplitude contrast. Take for example the topography (black curve) and the amplitude (red curve) line profiles corresponding to E=1.46 Vμm$^{-1}$. At the highest topographic point at ~5.2 μm lateral distance we observe the lowest amplitude signal indicative of an insulating state that resulted due to proton accumulation. The correlation between the topographic height changes and the corresponding near-field contrasts of H-NNO are clearly shown in Fig. S2. These changes induced by proton drift are plotted in reference to the initial state (i), shown in Fig. 1d. At all electric field values (E=1.12, 1.46, and 0 V/μm), changes in amplitude are consistently correlated with changes in height, indicating that the two mechanisms are concurrent in dopant drift. Pearson's correlation coefficient, which measures the covariance of the two variables divided by the product of their standard deviations, has been used to quantify this correlation. Positive values in Fig S2 confirm this correlation between height and amplitude. Moreover, at increasing fields, we note the right-drift of the stripes normal to the field, likely resulting from collective motion of hydrogen -rich regions. Inset in Fig. 1(f) shows the first derivative of the amplitude line profiles. With increasing E-field, the location on the horizontal axis where the first derivative becomes zero (amplitude minima) is shifted to the right (away from the left electrode) indicating that the heavily doped stripe parallel to the electrode edge is moving to the right.

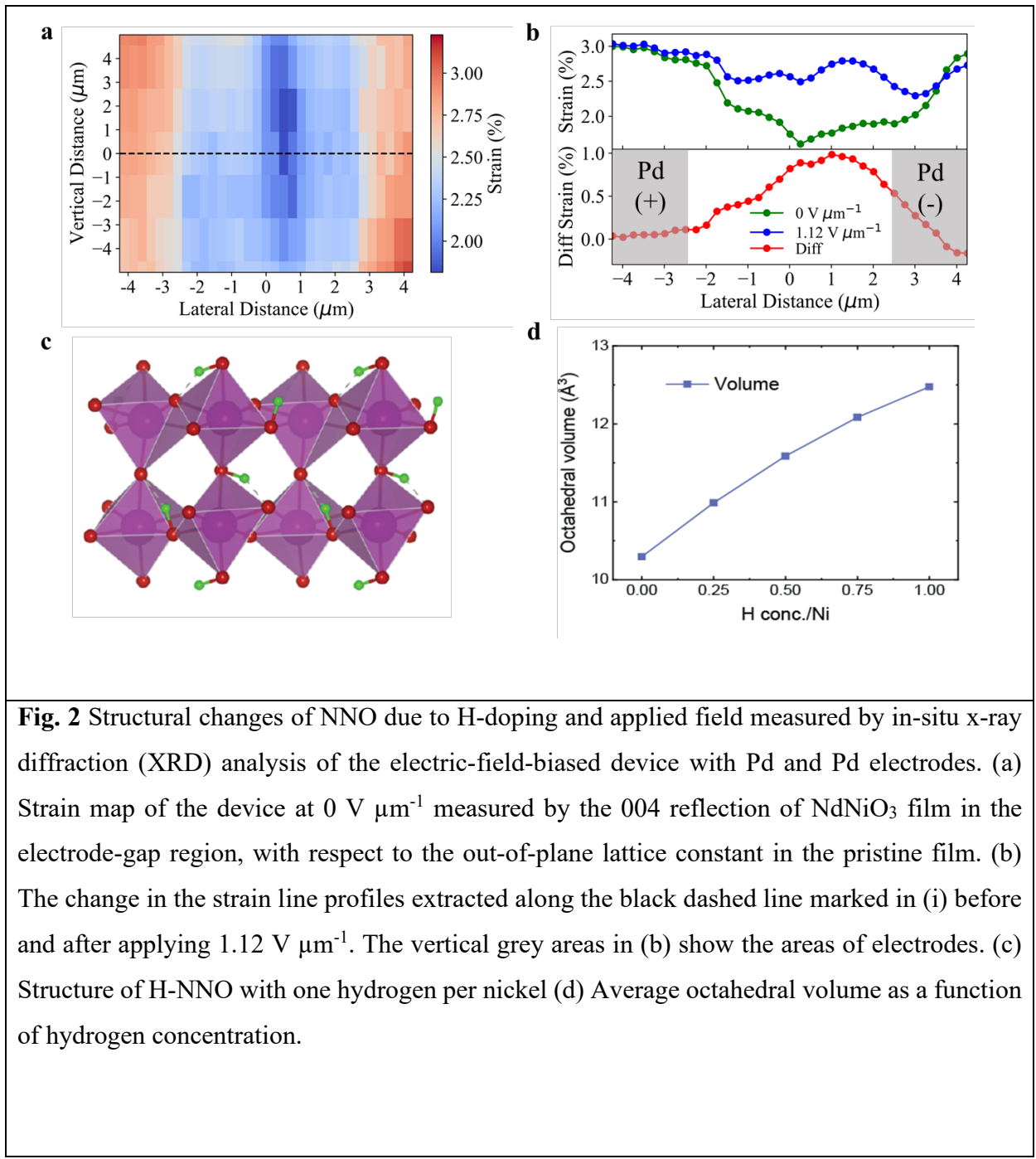

**Fig. 2** Structural changes of NNO due to H-doping and applied field measured by in-situ x-ray diffraction (XRD) analysis of the electric-field-biased device with Pd and Pd electrodes. (a) Strain map of the device at 0 V μm$^{-1}$ measured by the 004 reflection of NdNiO$_3$ film in the electrode-gap region, with respect to the out-of-plane lattice constant in the pristine film. (b) The change in the strain line profiles extracted along the black dashed line marked in (i) before and after applying 1.12 V μm$^{-1}$. The vertical grey areas in (b) show the areas of electrodes. (c) Structure of H-NNO with one hydrogen per nickel (d) Average octahedral volume as a function of hydrogen concentration.

To further explore how the structure of the NNO layer in these devices changes after H-doping, X-ray nano-diffraction imaging experiments were performed at the 7-ID-C beamline of the Advanced Photon Source (APS). The strain map across the gap (Figure 2a) was obtained by evaluating the measured lattice constant with respect to the lattice constant of the film outside the device far from the electrodes, which is close to that of the pristine NNO film.[12] The positive strain implies that the H-doping of the sample leads to an out-of-plane expansion of the NNO, a behavior that has also been seen in other doped materials.[19] This out-of-plane expansion should make the bonds weaker and softer, which is consistent with what was suggested by the s-SNOM measurements shown in Fig. 1d. The s-SNOM measurements have also shown that the higher the local H concentration is within the NNO film, the more the NNO film expands in the out-of-plane direction. Based on this, the larger strain values under the electrodes (red regions in Fig. 2a) imply that there is a higher H concentration under the electrodes compared to the H concentration in the gap of the device.

*In-situ* electrical biasing experiments were also performed to understand how the H-NNO crystal structure changes with DC bias. A series of diffraction images were collected along the black dashed line in Figure 2a to extract the strain profile across the gap before applying an E-field. A DC field of 1.12 V $\mu m^{-1}$ was then applied across the device gap, and after waiting for the current through the device to stabilize, the X-ray diffraction measurements were repeated in the same region across the same region in the gap. The differences of the strain lineouts are plotted in Fig. 2b. The strain inside the gap increases as a result of an increase in the out-of-plane lattice constant. This lattice expansion is consistent with a field-driven hydrogen transport process, although the magnitude of change in strain is smaller than what is observed in ionic liquid gating experiments.[12] Our observation unambiguously shows that lattice expansion correlates with the M/I phase changes revealed by s-SNOM when an E-field is applied (Figure 1d&e).

These observations are further supported by first-principles calculations. We investigated the effect of the hydrogen concentration on the volume expansion of $H_x$-NNO with $x$ values of 0, 0.25, 0.50, 0.75 and 1.00/Ni atom using Density Functional Theory (DFT) calculations. Figure 2b reveals a nearly linear relationship between the H dopant concentration in NNO and the expansion of $NiO_6$ octahedra, with changes of approximately 6.71%, 12.6%, 17.3%, and 21.2% for H per Ni atom ratios of 0.25, 0.5, 0.75, and 1.0, respectively. A corresponding optimized structure at one H per nickel is presented in Figure 2b. Upon H addition, we find that the electron from the hydrogen is transferred to a neighboring nickel atom. This results in a strong volume expansion of the nickel – oxygen octahedra as the hydrogen concentration increases. This excess charge localization process has been previously reported in related systems[20] and has been associated with a significant renormalization of the electronic structure, with H-NNO becoming a wide band gap insulator. Importantly, the charge localization results in a significant volume increase driven by the octahedral expansion, as shown in Fig. 2c. Interestingly, we find this volume expansion occurs independently of the metallic or insulating nature of the ground state in our simulations, suggesting electron localization and the associated volume expansion will occur in all of the hydrogen-rich regions of the device.

Beyond the structural and nanoscale conductivity changes described above, the effect of hydrogen dopants can, in principle, be probed by the local spectroscopic signatures associated with dopant vibrational states. However, thus far the real-space spectroscopic signature of these states has remained elusive. We use hyperspectral nano-FTIR imaging to map vibrational states of H-NNO and the dependence of these states on the dopant concentration. We employed an H-NNO sample with asymmetric electrodes (Pd-Au) since this configuration enables to create an asymmetric dopant concentration gradient across the region between the electrodes, with the highest concentration of dopant closest to the Pd electrode (for catalytic hydrogen dissociation) and decreasing away from it towards the Au electrode. This gradient is clearly visible in the near-field amplitude contrast image (Fig. 3a(ii)) and the 2D hyperspectral amplitude (Fig. 3b(i)) and phase maps (Fig. 3b(ii)).

The 2D hyperspectral phase and amplitude maps of the H-NNO sample were obtained by placing the tip of the microscope at 10 different equally spaced points along the yellow arrow line shown in the topography image in Fig. 3a(i). The images show strong intensity and spatial variation, particularly in the spectral range of 640-780 $cm^{-1}$. We extracted the phase spectral line profiles of each of the 10 points in the range of 640-1450 $cm^{-1}$ and plotted in Fig. 3c. We also show phase spectra of just 6 points in the range 1450-3000 $cm^{-1}$ in Fig. 3d, as the spectra for points 7-10 were found to be too noisy in this frequency range. The spectral range of 640-780 $cm^{-1}$ shows a clear dependence of intensity and peak position on location, which represents dopant concentration. The phase peak at point 1, closest to the Pd electrode, shows the strongest peak intensity and the signal intensity progressively decreases for points away from the Pd electrode.

The vibrational spectra of H-NNO shown in the spectral range 600-3000 $cm^{-1}$ are complicated by the presence of overlapping metal-oxygen and metastable vibrational modes that arise due to the hydrogen doping. To further understand the nature of these modes, we computed the phonon density of states using density functional perturbation theory calculations (DFPT) of H-NNO. As shown in Figure 2b, we consider the orthorhombic phase (*Pbnm*) of NNO with a concentration of hydrogen varying from 0.25 to 1 hydrogen per nickel in a 16 nickel atom periodic approximant. Complete computational details are given in the Methods section.

Computed zone-centered phonon frequencies and the corresponding decomposition of the vibrational eigenstates are given in Fig. 4a for one representative hydrogen concentration of 0.25H/Ni (the phonon density of states for hydrogen-free NNO and other hydrogen concentrations are shown in Fig. S3). Our calculations predict a nearly continuous phonon spectrum up to ~600 cm$^{-1}$. This prediction agrees with the results of the experimental measurement shown in Fig. 3c. The heavier vibrational modes associated with the Nd and Ni-dominated optical branches have frequencies up to 200 cm$^{-1}$ and 300 cm$^{-1}$ respectively, while the 300 cm$^{-1}$ to 600 cm$^{-1}$ frequency range is dominated by collective

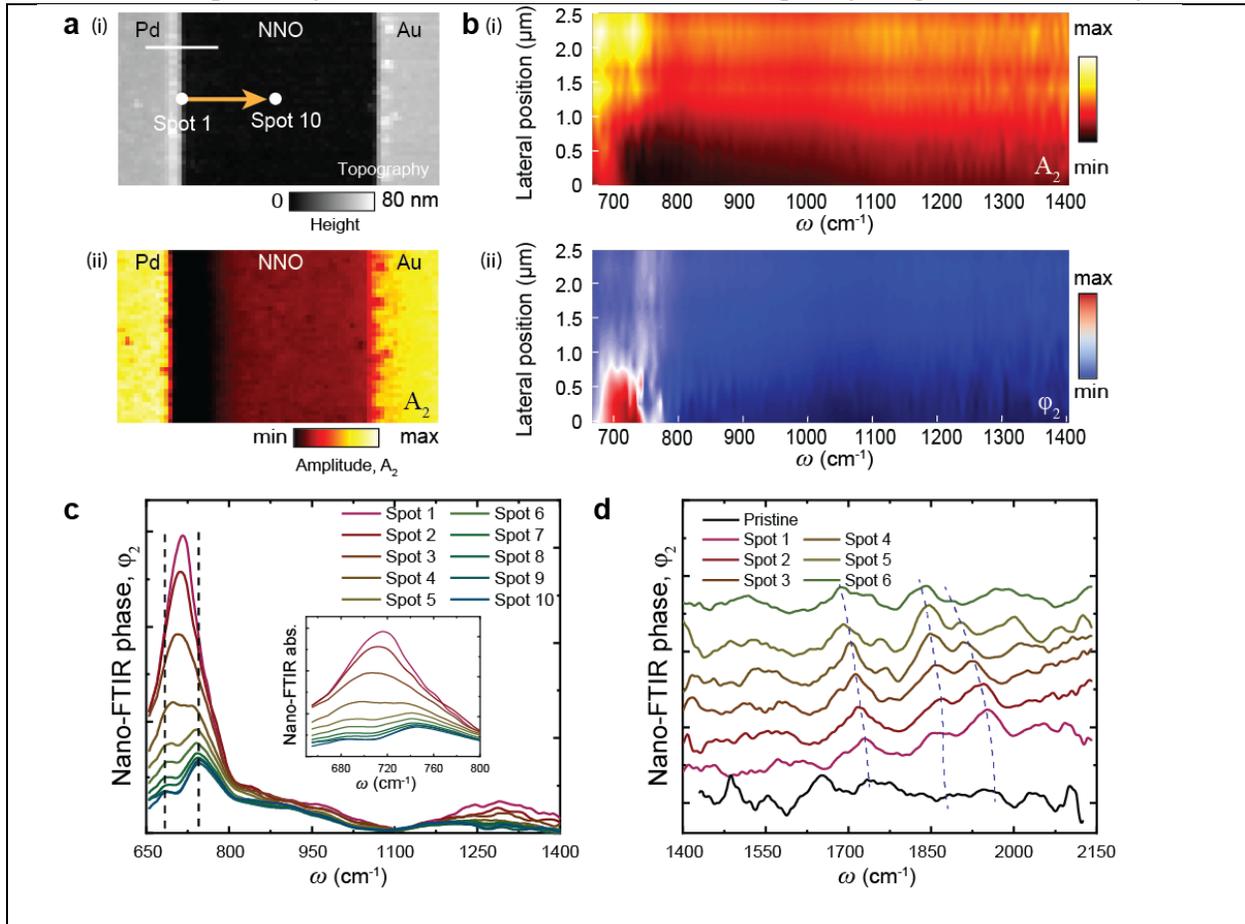

**Fig. 3** Nano-spectroscopy of H-NNO revealing spectral evolution of vibrational modes as a function of doping level. (a) (i) Topography and (ii) near-field amplitude images of H-NNO region between the Pd and Au electrodes, captured at λ= 10.5 μm. (b) Hyperspectral (i) amplitude and (ii) phase map obtained along the 2 μm length marked by yellow arrow in the topography image in (a). (c) Extracted phase spectra from the hyperspectral phase map in (b-(ii)) ranging from most doped (spot 1) to least doped (spot 10) regions. (d) Phase spectra extracted from spot 1-6 together with a spectrum from a pristine NNO (black line) and all spectra are normalized to that obtained on a Au reference surface, and are vertically offset for clarity. Scale bar in (a-(i)) is 2 μm.

motions of the oxygen octahedra (e.g., mode M$^1$ in Fig 4v). The M–O, O–M–O, and M–O–M (M =

Ni, Nd) stretching vibrations modes are also expected to be present in the experimental broad absorption band near 600 cm$^{-1}$ (Fig. 3e).

The addition of hydrogen results in two major changes in the vibration spectrum. First, OH stretching modes appear at higher frequencies, with strong variability of the frequency with respect to the hydrogen configuration and concentration (3000 cm$^{-1}$ for the longitudinal and $\sim$ 1064 cm$^{-1}$ for the transverse motions for the configuration in Fig 4b). Second, when placed near an oxygen, the hydrogen perturbs the octahedral motion resulting in high-energy modes of the unbound oxygens (see mode M$^2$ in Fig 4b) slightly above the oxygen vibrational continuum. The maximum frequency observed in pristine NNO is 596 cm$^{-1}$, while an oxygen dominated mode appears at 601 cm$^{-1}$ in the H$_{0.25}$NdNiO$_3$ system. While the frequencies associated with hydrogen motion are likely to strongly vary in this system with many metastable states[21], our calculations clearly indicate that vibrational modes with frequencies in excess of 700 cm$^{-1}$ are associated with OH stretching modes.[22-25]

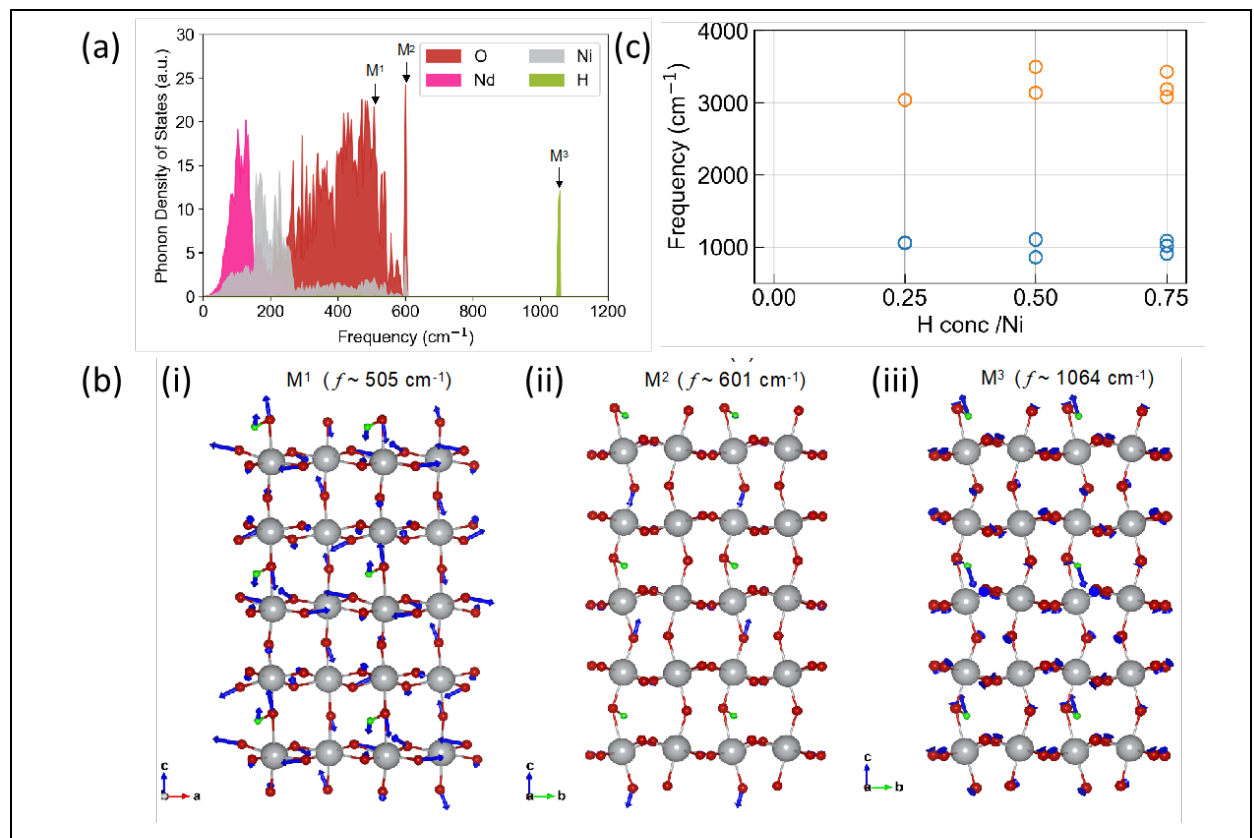

**Fig. 4** DFT analysis of H doping effects on the vibrational modes of NNO. (a) Phonon Density of States for 0.25H-NNO. (b) Phonon eigenvectors (depicted by blue arrows) illustrating modes with frequencies: (i) 505 cm$^{-1}$, associated with the collective motion of NiO$_6$ octahedra, (ii) 601 cm$^{-1}$, indicating the vibration of O atoms, and (iii) 1064 cm$^{-1}$, reflecting the vibration of H atoms.

Additionally, (c) demonstrates the observed vibrational hybridization and spectral broadening upon the addition of H.

These calculations agree with experimental observations of increased intensities of the broad peaks in the frequency range up to around 800 cm$^{-1}$ with H dopant concentration. The ionic attachment of hydrogens to oxygen means that several combinations of O-H stretch, and intercalated hydroxyl groups can be generated in the range 1450-3000 cm$^{-1}$ as shown in Fig. 3d. Even in the absence of explicit configurational disorder, we find that $x$>0.25 concentrations lead to significant vibrational hybridization and spectral broadening. In particular, we find that hybridization of the OH modes of neighboring octahedra results in vibrational modes splitting on the scale of 100 to 350 cm$^{-1}$. These modes appear to shift to lower energy with increasing dopant concentration. The spectral range 1600-1950 cm$^{-1}$ is expected to be dominated by dopant related stretching and bending modes, and strikingly the peaks show a general trend of red shift as the concentration of the dopant increases (moving from Spot 10 to Spot 1). This is expected since strain increases with dopant concentration, which results in a spectral red shift and confirms dopant-based vibrations, as predicted by DFPT.

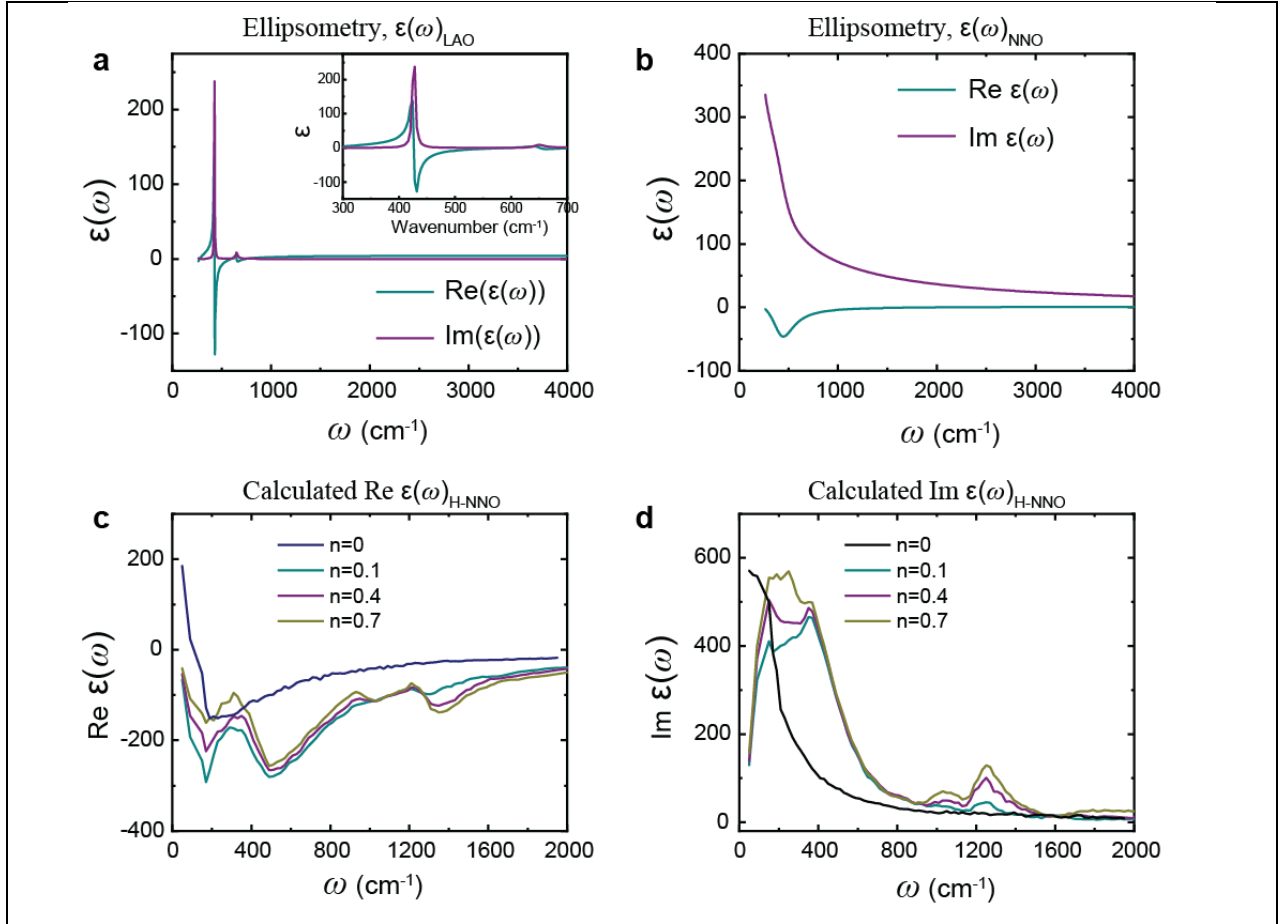

**Fig. 5** Experimental far-field ellipsometry real (green) and imaginary (purple) parts of the dielectric function of LAO (a) and pristine NNO on LAO substrate (b) in the frequency range 400-4000 cm$^{-1}$. Inset in (a) shows the zoomed in region from 300-700 cm$^{-1}$. Theoretical calculations of (c) real and (d) imaginary dielectric function of NNO at different doping level, n = 0, 0.1, 0.4, and 0.7. The simulations were done at a dopant interaction strength, $\epsilon_{LJ}$=1.54×10$^{-23}$ and σ=2×10$^{-10}$.

The response of H-NNO to interaction with light is characterized by the complex-valued local dielectric function $(\varepsilon(\omega))$ of the sample. The near-field vibrational absorption spectra $(\varphi_n)$ presented in Fig. 3 can be directly assigned based on the imaginary part of the dielectric function (Im $\varepsilon(\omega)$) of HNNO. This is due to the direct proportionality of the near-field phase $(\varphi_n)$ to the Im $\varepsilon(\omega)$ for weak oscillatory modes, such as the dopant induced vibrational modes we analyze in this work. As such, mapping the dielectric function of HNNO will provide direct information on the origin and assignment of the vibrational modes shown in Fig. 3. To that end we performed ellipsometry measurements in the frequency range from 400 cm$^{-1}$ to 4000 cm$^{-1}$, using an IR-VASE ellipsometer (J. A. Woollam Co., Lincoln, NE) of pristine NNO on LAO substrate and the pristine LAO substrate. Samples were mounted on a precision rotation stage, and a data set was acquired at a spectral resolution of 8 cm$^{-1}$. After the measurement, the data was fit using standard numerical analysis methods (similar to Jellison[26,27] and also Herzinger[28]) using the LAO optical function for the substrate. The real Re($\varepsilon(\omega)$) and imaginary Im($\varepsilon(\omega)$) parts of the dielectric function ($\varepsilon(\omega)$) of NNO were calculated with a resistivity-scattering time Drude model[29] plus a Lorentz oscillator to account for the additional absorption centered around 430 cm$^{-1}$. The WVASE program from J.A. Woollam Co. was used to build the model and fit the data. Figure 5a-b show the real (black solid line) and imaginary parts (red solid line) of the dielectric functions extracted from ellipsometry, for the LAO substrate and the NNO/LAO sample, respectively.

We note, however, that it is extremely challenging to perform ellipsometry measurements on NNO (H-NNO) with various dopant levels and to quantify the nanoscale changes of $\varepsilon(\omega)$ due to both sensitivity and resolution issues. To circumvent this difficulty, we implemented a simulational methodology that we recently developed to acquire information about the nanoscale $\varepsilon(\omega)$ of H-NNO.[30] Using our model, we first reproduced the undoped dielectric function of NNO acquired from ellipsometry measurement (Fig. 5c-d) and then predicted changes to its response at different dopant levels. This method considers harmonic interactions between bonded atoms and uses a combination of Langevin dynamics and Monte Carlo methods to investigate the frequency dependent dielectric function of a sample (see Ref[30] for details). The undoped dielectric function of NNO was reproduced by fitting appropriate harmonic coupling constants as well as damping parameters to correspond to experimental results (Fig. 3c-d). To model the effects of dopants in the system we randomly placed particles interstitially throughout the lattice, where they interact with surrounding particles via a Lennard-Jones potential (see Methods section for details). The dopant level is controlled with a parameter $n \in [0,1]$ which describes the number of dopants per unit cell in the system.

We vary both the interaction parameters as well as the dopant concentration to explore broad feature modulations seen in the resultant calculated dielectric function. We found that increasing the interaction strength between the dopants and the lattice produces a new peak in the dielectric function (Im($\varepsilon(\omega)$) in Fig. 5(c-d) that increase in intensity as the interaction strength increases, which is similar to what is observed in the experiment (Fig. 3). When we allow the dopants in the system to couple to the lattice in a more complex way by including bond-angle and bond-length interactions with the surrounding crystal ions, we see multiple new peaks emerging. This indicates that the origin of the vibrational peaks is directly related to changes in Im($\varepsilon(\omega)$) due to both the dopant level (n) and the changes in corresponding coupling strengths of the dopants. These 3-body potentials also correspond to the bond bending and stretching modes described above in the DFPT calculations (Fig. 4), reinforcing this interpretation. We also leverage Monte Carlo techniques to investigate structural distortion induced by the inclusion of these dopants in our model. We allow the dopant particles to interact with the ions and allow the system to relax into its minimum free-energy configuration. Once we have established that the system has sufficiently relaxed, we calculate the total volume change of the system and perform such calculations over a variety of interaction parameters as well as dopant levels to ascertain overall trends. We find that the increase in dopant concentration leads to an overall increase of average unit cell volume, an effect which corresponds to the DFT calculation as well as the X-ray and s-SNOM measurements that we have described above in Fig. 3. Furthermore, using the dielectric values obtained from abovementioned calculations, we generated the near-field amplitude and phase spectra using an extended finite-dipole model[31] for NNO samples with different doping levels and the results are presented in Fig. S4 in the supplemental section.

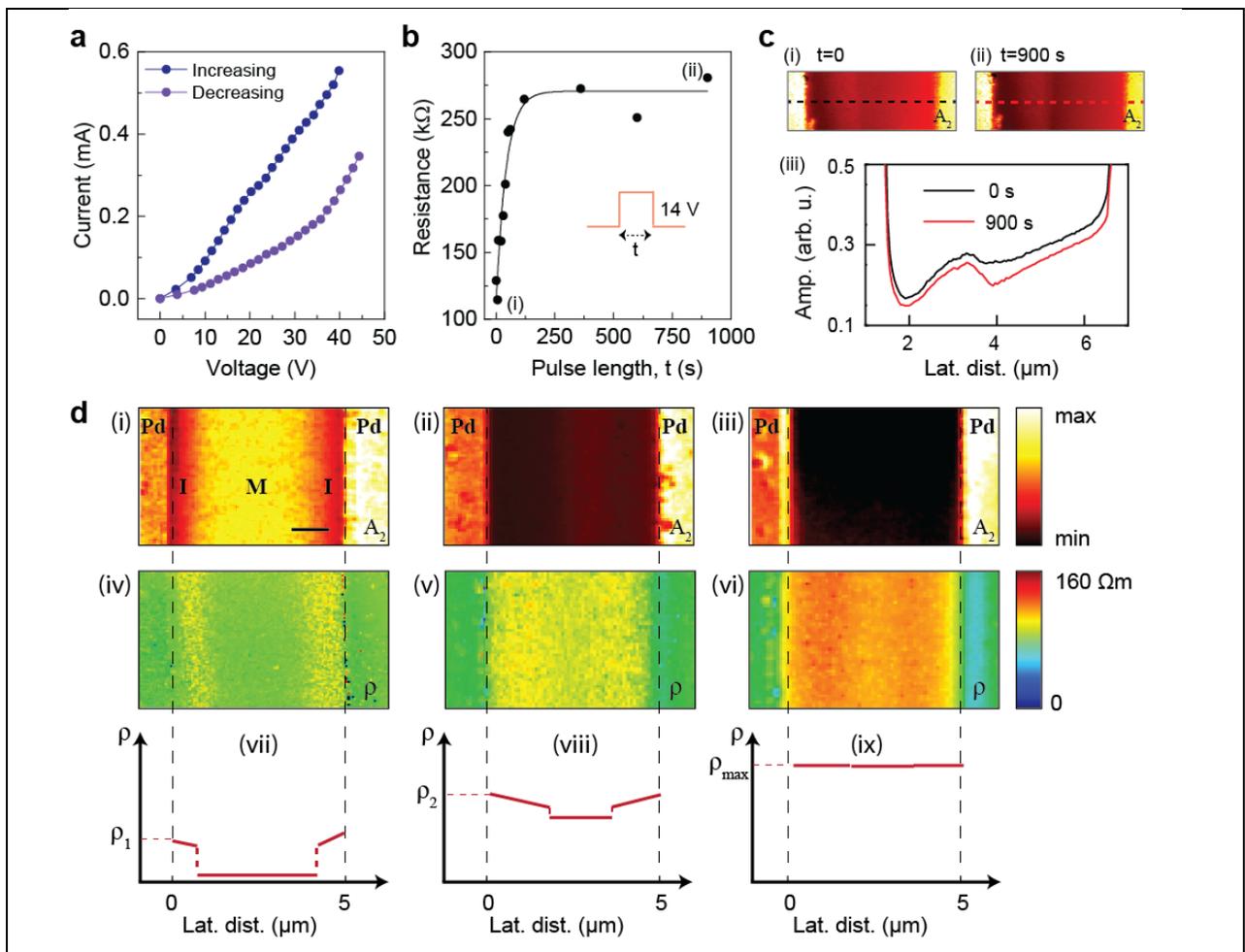

**Fig. 6** Effect of doping on global and local conductivity properties H-NNO. (a) Current vs voltage relationship of H-NNO device with increasing (blue dots) and decreasing (purple dots) voltage across the device. (b) Global resistance change across the device with voltage pulse duration (black dots), black solid line is guide to the eye. (c) Amplitude images taken at t=0 s (i), t=900 s (ii) also shown in (b), and the line profiles extracted from those two images (iii) along the horizontal dashed lines marked in c(i) and c(ii). (d) Near-field 2nd harmonics amplitude images of (i) lightly doped (40 Ω), (ii) medium doped (125 kΩ) and (iii) heavily doped (430 kΩ) global resistance states of the device. All near-field images were obtained at laser illumination wavelength of λ= 10.5 μm. The near-field amplitude images in (i-iii) are converted to resistivity maps shown in (iv-vi). For this conversion, the $Im(\varepsilon)$ values were obtained from calculations shown in (c-d) for H-NNO with different dopant levels. The plots (vii-ix) are resistivity profile images in (iv-vi). Scale bar in d(i) is 1 μm.

The device's global (between electrodes) forward and reverse current (I) and voltage (V) relationship for V ranging between 0 to 45 V shown in Fig. 6a. resembles the hysteresis pattern reported for similar synaptic devices.[20] The dependence of global resistance on the pulse length is shown in Fig. 6b. With increasing pulse length, global resistance exponentially increases and saturates at a level ~275 kΩ after the pulse length increased beyond 200 s. We display near-field amplitude images of two pulse lengths in Fig. 6c(i) (t=0 s) and 6c(ii) (t=900 s) and the corresponding line profiles extracted along the horizontal dashed lines are shown in Fig. 6c(iii). After electrical pulsing, the change in the resistance distribution across the gap is demonstrated by the changes in the line profiles at t=0 and t=900 s. The decrease in amplitude signal is captured by the line profiles for t=900 s which indicates an increase in resistance after the applied pulses. For experiments presented in Fig. 6b, the voltage pulse magnitude was kept at 14 V.

IR near-field images can be converted to resistivity maps providing nanoscale electronic properties of the oxide. To that end we imaged the Pd-NNO-Pd device at three different resistance levels that were produced by hydrogenation and application of pulsed E-fields between the electrodes. Figure 6d shows three subplots (i-iii) which show near-field s-SNOM amplitude images of H-NNO at three different resistance states: 40 Ω, 125 kΩ, and 430 kΩ. The images were obtained using a laser excitation wavelength of 10.5 μm. The near-field images were then converted to resistivity maps shown in Fig. 6d(iv-vi). This is achieved by calculating the optical conductivity, $\sigma_{(RE)}$ of the sample at each pixel using the equation $\sigma_{(RE)} = \frac{\omega}{4\pi} \varepsilon_{(IM)}$ where $\varepsilon_{(IM)}$ is the imaginary part of the dielectric function of the sample[32-34]. To quantitively extract the value of $\varepsilon_{(IM)}$ at each pixel of the near-field image we relied on our simulational method described above which provides the dielectric function of doped NNO at different dopant levels (see Methods section for details). The resulting resistivity maps (Fig. 6e(iv-vi)) show similar contrast profile as the near-field images and provide qualitative conducting states of the sample at a pixel level.

**Conclusion**

Nanoscale infrared imaging, in-situ high-resolution x-ray diffraction and spectroscopy measurements reveal simultaneous structural and local conductivity modulation in H-NdNiO₃ incurred by hydrogen dopants that can be actively controlled by applying electric fields. First-principles calculations, dielectric function simulations, and real-space infrared hyperspectral nanoimaging unveil a range of vibrational modes that result from hydrogen doping. The results provide insight into electronic phase transitions mediated by ions at a length scale otherwise inaccessible in operando devices, and how different dopants affect their electronic properties.

**Methods**

*Sample preparation*

NNO films were deposited on 400 um thick lanthanum aluminate (LaAlO₃) substrates by physical vapor deposition (PVD) sputtering technique. The thickness of the NNO films investigated in this study was 60 nm. For electrical contacts, Pd-Pd or Pd-Au electrode pads (100 µm x 100 µm) of thickness 80 nm were deposited on NNO films using electron beam lithography. Fabricated NNO devices were hydrogenated in a cell in a forming gas flow (hydrogen/nitrogen 5%) at 120 °C for ~20 minutes while measuring the resistance between two Pd electrodes using an ohmmeter, until the measured resistance reaches the desired level.

*Nanoimaging and nanospectroscopy*

A commercial (Neaspec GmbH) scattering type scanning near-field microscope (s-SNOM) which is based on a tapping mode AFM was utilized to perform the near-field imaging of the H-NNO devices. A metal coated cantilevered AFM tip oscillating at a frequency of $\Omega$~ 280 kHz with a tapping amplitude of ~100 nm was used to obtain both topography and near-field amplitude images. Spatial resolution is only limited by the tip-apex radius independent of the wavelength from visible to terahertz spectral range[31,35,36]. For this work for nanoimaging we have used a monochromatic quantum cascade laser of frequency 943 cm⁻¹ and for nanospectroscopy a broadband light source in the spectral range of 600-2100 cm⁻¹. The incident beam is focused on the tip-sample interface by a parabolic mirror at angle of 45° to the sample surface and backscattered light from the interface was detected and demodulated at higher harmonics of tip resonance frequency using phase modulation interferometry (for single frequency imaging) and asymmetric Fourier transform Michelson interferometer (for nanospectroscopy).

*X-ray nanodiffraction imaging*

The x-ray nanodiffraction measurements were performed at beamline 7ID-C of the Advanced Photon Source[37]. The 11.5 keV x-ray beam is focused by a Fresnel zone plate to a vertical spot size of 400 nm and a horizontal spot size of 1.2 μm full width at half maximum (FWHM). The device sample was mounted vertically for a horizontal diffraction geometry, so the direction of the applied electric field was parallel to the vertical direction and the long dimension of the gap of the device was aligned along the horizontal direction. The sample was raster scanned with the x-rays along the in-sample plane directions, and the diffraction patterns were collected at various points in real space on the sample. Due to the broad rocking curve of the H-NNO film (Fig. S4), the H-NNO diffraction peak across a large range of reciprocal space along the out-of-plane reciprocal lattice can be monitored at an incident angle of 15.7°. The corresponding average lattice constant can be derived from the 2θ angle of the diffraction peaks. The centroid of the diffraction peak was measured by fitting the H-NNO x-ray diffraction peak on the detector along the 2θ direction. The DC bias was applied by a Keithley 2612 Source meter to the electrodes. The device reaches a steady state in about 10 minutes, which was confirmed by monitoring the current across the electrodes.

*Density Functional Theory (DFT) calculations*

All the density functional theory calculations were performed using the Vienna ab initio Simulation (VASP)[38] package with Perdew-Burke-Ernzerhof[39] Generalized Gradient Approximation (GGA)[40] exchange correlation functional. The rotationally invariant form of GGA + U from Ref[41,42] with U = 4.6 eV and J = 0.6 were used to treat the strong Coulomb repulsion among the Ni 3d electrons for NNO, where U is the on-site Coulomb parameter. All calculations used 600 eV as the plane wave cutoff energy. The total energies' electronic convergence criteria were set at $10^{-6}$ eV. The structures were optimized using a conjugate gradient approximation[43] as implemented in VASP until all the atomic forces were < 0.01 eV/Å. The spin-polarized calculations were performed at the Gamma points using a k-point density of 1000, where the k-point density was determined by multiplying the number of atoms in the simulation cell ($n_{atoms}$) by the number of k-points ($n_{kpoints}$) [44]. The symmetry was turned off in all calculations. The metallic phases of NNO were considered with ferromagnetic initializations whereas in case of insulation phase we have considered the anti-ferromagnetic settings with T type antiferromagnetic ordering[45]. Mixing parameters such as AMIX = 0.1 and BMIX=0.001 was used to expedite the electronic convergence. Phonon modes and frequencies of the optimized structures at different level of H concentration for both the phases was calculated using finite differences approaches.

*Molecular Dynamics (MD) calculations*

We simulate the dielectric response of pristine NNO using a Langevin dynamics methodology described by Hancock et. al.[30], which models condensed matter systems as networks of damped oscillators with additional couplings to an oscillating external E-field. Systems simulated consisted of 5 coupled 20×20 arrays of atoms, each with periodic boundary conditions, all fixed at constant temperature. Typical timesteps for integration were 10 femtoseconds and the total integration time was 100 picoseconds. To obtain precise mean values and calculate statistical error bars, 50 independent simulations were performed for each set of conditions. We set the resonant frequency of the $Nd - O$ bond $\omega_{0,Nd-O} = 150 \ cm^{-1}$. Similarly, we set $\omega_{Ni-O} = 125 \ cm^{-1}$. Damping parameter $b/m_e$, where $m_e$ is the mass of the bond charges in the simulation, was set to $\frac{b}{m_e} = 2 \times 10^2 \ \frac{Ns}{m \ kg}$. We found that such a modelling scheme would be appropriately fast and flexible to explore different interactions and their resultant effects on the computed dielectric function. To fully investigate the role of $H^+$ dopants in dielectric modulation of NNO, we require an appropriate model for the dopant-lattice interactions. As such, we placed dopants randomly throughout the lattice interstitially at some predetermined concentration $n$ and assume for simplicity that such particles are static throughout the duration of the simulation. These dopants interact with surrounding lattice ions via a Lennard-Jones potential given by:

$$V_{LJ} = 4\epsilon \left( \left(\frac{\sigma}{r}\right)^{12} - \left(\frac{\sigma}{r}\right)^6 \right)$$

where $\epsilon$ describes the overall strength of the dopant interaction, $\sigma$ similarly describes the spatial extent of the interaction, and $r$ is the interparticle distance.

Different choices of the dopant interaction strength $\epsilon$ shifts the associated peak position in frequency (Fig 3 (c)), which bolsters this interpretation. Furthermore, an increase in the dopant concentration $n$ increases the prominence of these dopant induced resonances shown in Fig 3 (d), as one would expect. This computational procedure has allowed us to gain helpful insight into the manner by which dopants can impact the dielectric response of NNO in a way comparable with experiments. We then endeavored to increase the sophistication of our dopant modelling one step further by relaxing the static dopant restriction and adding in bond angle interactions between the dopants and the lattice. We, thus, introduce the three-body potential:

$$V_\theta = \frac{k_{ijk}}{2} \left( \cos \theta_{ijk} - \cos \theta_0 \right)^2$$

where $k_{ijk}$ is the associated three-body interaction strength between a dopant $i$ and particles $j$, $k$, $\theta_{ijk}$ is their instantaneous bond-angle, and $\theta_0$ is their bond angle at dopant equilibrium. Therefore, we have a combined dopant interaction potential represented by:

$$V_{dop} = V_{LJ} + V_{\theta}$$

As explained above, we scan through different selections of $k_{ijk}$ to show how the introduction of a bond-angle interaction might change the resonance landscape seen in the dielectric function. In Fig 3e, we have represented two spectra, one when $k_{ijk} = 0$, and one where $k_{ijk} \neq 0$. In both cases, we kept $\epsilon$ the same to better illuminate the role of the bond-angle interactions on producing more complex dielectric response as seen in experiments. We see the emergence of three dopant peaks when $k_{ijk} \neq 0$ and only one otherwise. The addition of these three-body interaction allows for more complex interactions in the system, which is represented by the more numerous peaks in the spectra. This indicates that these interactions drive much of the unique and surprising features seen in the optical response of $H^+$ doped $NNO$.

**Acknowledgement**

Support for S.G., S.S., M.G., and Y.A. is provided by the Air Force Office of Scientific Research (AFOSR) grant number FA9550-19-0252. Partial support for S.G. comes from the National Science Foundation (NSF) Grant No. 2152159 (NRT-QuaNTRASE). The sample fabrication was supported as part of the Quantum Materials for Energy Efficient Neuromorphic Computing (Q-MEEN-C), an Energy Frontier Research Center funded by the U.S. Department of Energy (DOE), Office of Science, Basic Energy Sciences (BES), under Award # DE-SC0019273. This work was also supported in part by NSF Grant #1904097 and in part by resources and technical expertise from the Georgia Advanced Computing Resource Center, a partnership between the University of Georgia's Office of the Vice President for Research and Office of the Vice President for Information Technology. M.Z. and H.W. acknowledge the support for X-ray diffraction measurements by the U.S. Department of Energy, Office of Science, Basic Energy Sciences, Materials Sciences and Engineering Division. Work performed at the Center for Nanoscale Materials, a U.S. Department of Energy Office of Science User Facility, was supported by the U.S. DOE, Office of Basic Energy Sciences, under Contract No. DE-AC02-06CH11357.

**Author contributions**

Y.A. and SR conceived the idea and guided the overall project. Q.W., T.J.P. and S.R. deposited the NNO films and fabricated the devices. S.G., K.Y and Q.W. designed the NNO devices. S.G., S.S

M.G. performed the near-field imaging and spectroscopic experiments and analyzed the data. S.M., P.D. and S.S. performed the DFT calculations. S.H. and D.P.L. performed the M.D. and MCS calculations. M.Z. and H.W. performed the nano-XRD experiments. SG and YA wrote the manuscript with input from all authors. All authors read and approved the final manuscript.

**Competing interests**

The authors declare no competing interests.

**Materials & Correspondence**

Data will be made available upon reasonable request. All correspondence and material requests should be addressed to yohannes.abate@uga.edu.

**References**

1    Park, T. J. *et al.* Complex Oxides for Brain-Inspired Computing: A Review. *Adv Mater*, doi:10.1002/adma.202203352 (2022).

2    del Valle, J., Ramirez, J. G., Rozenberg, M. J. & Schuller, I. K. Challenges in materials and devices for resistive-switching-based neuromorphic computing. *Journal of Applied Physics* **124**, doi:10.1063/1.5047800 (2018).

3    Schuller, A. *et al.* Neuromorphic computing: Challenges from quantum materials to emergent connectivity. *Appl Phys Lett* **120**, doi:Artn 14040110.1063/5.0092382 (2022).

4    Markovic, D., Mizrahi, A., Querlioz, D. & Grollier, J. Physics for neuromorphic computing. *Nat Rev Phys* **2**, 499-510, doi:10.1038/s42254-020-0208-2 (2020).

5    Boybat, I. *et al.* Neuromorphic computing with multi-memristive synapses. *Nature Communications* **9**, doi:10.1038/s41467-018-04933-y (2018).

6    Catalano, S. *et al.* Rare-earth nickelates RNiO3: thin films and heterostructures. *Reports on Progress in Physics* **81**, doi:10.1088/1361-6633/aaa37a (2018).

7    Shi, J., Ha, S. D., Zhou, Y., Schoofs, F. & Ramanathan, S. A correlated nickelate synaptic transistor. *Nature Communications* **4**, doi:10.1038/ncomms3676 (2013).

8    Li, J. R., Ramanathan, S. & Comin, R. Carrier Doping Physics of Rare Earth Perovskite Nickelates RENiO3. *Front Phys-Lausanne* **10**, doi:ARTN 834882
10.3389/fphy.2022.834882 (2022).

9       Zhang, H. T. *et al.* Perovskite neural trees. *Nat Commun* **11**, 2245, doi:10.1038/s41467-020-16105-y (2020).

10      Sidik, U., Hattori, A. N., Rakshit, R., Ramanathan, S. & Tanaka, H. Catalytic Hydrogen Doping of NdNiO3 Thin Films under Electric Fields. *Acs Applied Materials & Interfaces* **12**, 54955-54962 (2020).

11      Oh, C., Heo, S., Jang, H. M. & Son, J. Correlated memory resistor in epitaxial NdNiO3 heterostructures with asymmetrical proton concentration. *Appl Phys Lett* **108**, doi:Artn 12210610.1063/1.4944842 (2016).

12      Chen, H. W. *et al.* Protonation-Induced Colossal Chemical Expansion and Property Tuning in NdNiO3 Revealed by Proton Concentration Gradient Thin Films. *Nano Lett*, doi:10.1021/acs.nanolett.2c03229 (2022).

13      Bisht, R. S. *et al.* Spatial Interactions in Hydrogenated Perovskite Nickelate Synaptic Networks. *Nano Lett*, doi:10.1021/acs.nanolett.3c02076 (2023).

14      Preziosi, D. *et al.* Direct Mapping of Phase Separation across the Metal-Insulator Transition of NdNiO3. *Nano Lett* **18**, 2226-2232, doi:10.1021/acs.nanolett.7b04728 (2018).

15      Pereda, A. E. Electrical synapses and their functional interactions with chemical synapses. *Nat Rev Neurosci* **15**, 250-263, doi:10.1038/nrn3708 (2014).

16      Kwon, D. H. *et al.* Unraveling the Origin and Mechanism of Nanofilament Formation in Polycrystalline SrTiO3 Resistive Switching Memories. *Adv Mater* **31**, doi:ARTN 1901322 10.1002/adma.201901322 (2019).

17      Cheng, S. B. *et al.* Operando characterization of conductive filaments during resistive switching in Mott Vo(2). *P Natl Acad Sci USA* **118**, doi:ARTN e2013676118
10.1073/pnas.2013676118 (2021).

18      Zhang, K. N., Ganesh, P. & Cao, Y. Deterministic Conductive Filament Formation and Evolution for Improved Switching Uniformity in Embedded Metal-Oxide-Based Memristors?A Phase-Field Study. *Acs Applied Materials & Interfaces* **15**, 21219-21227, doi:10.1021/acsami.3c00371 (2023).

19      Muscher, P. K. *et al.* Highly Efficient Uniaxial In-Plane Stretching of a 2D Material via Ion Insertion. *Adv Mater* **33**, doi:ARTN 2101875
10.1002/adma.202101875 (2021).

20      Zhang, H. T. *et al.* Reconfigurable perovskite nickelate electronics for artificial intelligence. *Science* **375**, 533-+, doi:10.1126/science.abj7943 (2022).

21    Park, T. J. *et al.* Efficient Probabilistic Computing with Stochastic Perovskite Nickelates. *Nano Lett*, doi:10.1021/acs.nanolett.2c03223 (2022).

22    Ignatius Arockiam, S., Peter Pascal Regis, A. & John Berchmans, L. Synthesis and Characterisation of Nano crystalline Neodymium Nickelate (NdNiO$_3$) Powders using Low Temperature Molten Salt Technique. *Research Journal of Chemical Sciences* **2**, 6 (2012).

23    Maulana, M. I., Putri, B. D. A., Anggaini, R. P., Yuliani, H. & Suhendra, N. A New Synthesis and Characterization of NdNiO3 Perovskite Nanoparticles via Coprecipitation Method. *Chemistry Research Journal* **5**, 11 (2020).

24    Sivakumar, M., Pandi, K., Chen, S. M., Cheng, Y. H. & Sakthivel, M. Facile synthesis of perovskite-type NdNiO3 nanoparticles for an effective electrochemical non-enzymatic glucose biosensor. *New J Chem* **41**, 11201-11207, doi:10.1039/c7nj02156a (2017).

25    Fernandes, J. D. G. *et al.* Low-temperature synthesis of single-phase crystalline LaNiO3 perovskite via Pechini method. *Mater Lett* **53**, 122-125 (2002).

26    Jellison, G. E. Data-Analysis for Spectroscopic Ellipsometry. *Thin Solid Films* **234**, 416-422, doi:Doi 10.1016/0040-6090(93)90298-4 (1993).

27    Jellison, G. E. in *Handbook of Ellipsometry*   Ch. 3, (William Andrew, Inc., 2005).

28    Herzinger, C. M., Snyder, P. G., Johs, B. & Woollam, J. A. Inp Optical-Constants between 0.75 and 5.0 Ev Determined by Variable-Angle Spectroscopic Ellipsometry. *J Appl Phys* **77**, 1715-1724, doi:Doi 10.1063/1.358864 (1995).

29    Tiwald, T. E., Thompson, D. W., Woollam, J. A., Paulson, W. & Hance, R. Application of IR variable angle spectroscopic ellipsometry to the determination of free carrier concentration depth profiles. *Thin Solid Films* **313**, 661-666, doi:Doi 10.1016/S0040-6090(97)00973-5 (1998).

30    Hancock, S. B., Landau, D. P., Aghamiri, N. A. & Abate, Y. Langevin dynamics/Monte Carlo simulations method for calculating nanoscale dielectric functions of materials. *Phys Rev Mater* **6**, doi:ARTN 076001
10.1103/PhysRevMaterials.6.076001 (2022).

31    Abate, Y. *et al.* Nanoscopy of Phase Separation in InxGa1-xN Alloys. *Acs Applied Materials & Interfaces* **8**, 23160-23166, doi:10.1021/acsami.6b06766 (2016).

32    Aly, K. A. Comment on the relationship between electrical and optical conductivity used in several recent papers published in the journal of materials science: materials in electronics. *J Mater Sci-Mater El* **33**, 2889-2898, doi:10.1007/s10854-021-07496-9 (2022).

33    Qazilbash, M. M. *et al.* Mott transition in VO2 revealed by infrared spectroscopy and nano-imaging. *Science* **318**, 1750-1753, doi:10.1126/science.1150124 (2007).

34    Aghamiri, N. A. *et al.* Reconfigurable hyperbolic polaritonics with correlated oxide metasurfaces. *Nature Communications* **13**, doi:ARTN 451110.1038/s41467-022-32287-z (2022).

35    Chen, X. Z. *et al.* Modern Scattering-Type Scanning Near-Field Optical Microscopy for Advanced Material Research (vol 31, 1804774, 2019). *Advanced Materials* **34**, doi:ARTN 2205636

10.1002/adma.202205636 (2022).

36    Ogawa, Y., Minami, F., Abate, Y. & Leone, S. R. Nanometer-scale dielectric constant of Ge quantum dots using apertureless near-field scanning optical microscopy. *Applied Physics Letters* **96**, doi:Artn 063107

10.1063/1.3309692 (2010).

37    Zhu, Y. *et al.* Structural imaging of nanoscale phonon transport in ferroelectrics excited by metamaterial-enhanced terahertz fields. *Phys Rev Mater* **1**, doi:ARTN 060601

10.1103/PhysRevMaterials.1.060601 (2017).

38    Kresse, G. & Furthmuller, J. Efficient iterative schemes for ab initio total-energy calculations using a plane-wave basis set. *Phys Rev B* **54**, 11169-11186, doi:DOI 10.1103/PhysRevB.54.11169 (1996).

39    Perdew, J. P., Burke, K. & Ernzerhof, M. Generalized gradient approximation made simple. *Physical Review Letters* **77**, 3865-3868, doi:10.1103/PhysRevLett.77.3865 (1996).

40    Blochl, P. E. PROJECTOR AUGMENTED-WAVE METHOD. *Physical Review B* **50**, 17953-17979, doi:10.1103/PhysRevB.50.17953 (1994).

41    Liechtenstein, A. I., Anisimov, V. I. & Zaanen, J. Density-Functional Theory and Strong-Interactions - Orbital Ordering in Mott-Hubbard Insulators. *Phys Rev B* **52**, R5467-R5470, doi:DOI 10.1103/PhysRevB.52.R5467 (1995).

42    Zhang, Z. *et al.* Perovskite nickelates as electric-field sensors in salt water. *Nature* **553**, 68-+, doi:10.1038/nature25008 (2018).

43    Stich, I., Car, R., Parrinello, M. & Baroni, S. Conjugate-Gradient Minimization of the Energy Functional - a New Method for Electronic-Structure Calculation. *Phys Rev B* **39**, 4997-5004, doi:DOI 10.1103/PhysRevB.39.4997 (1989).

44      Manna, S., Gorai, P., Brennecka, G. L., Ciobanu, C. V. & Stevanovic, V. Large piezoelectric response of van der Waals layered solids. *J Mater Chem C* **6**, 11035-11044, doi:10.1039/c8tc02560f (2018).

45      Haule, K. & Pascut, G. L. Mott Transition and Magnetism in Rare Earth Nickelates and its Fingerprint on the X-ray Scattering. *Sci Rep-Uk* **7**, doi:ARTN 1037510.1038/s41598-017-10374-2 (2017).

**Supplementary Material**

**Infrared Nanoimaging of Hydrogenated Perovskite Nickelate Synaptic Devices**

Sampath Gamage[1], Sukriti Manna[2,3], Marc Zajac[4], Steven Hancock[5], Qi Wang[6], Sarabpreet Singh[1], Mahdi Ghafariasl[1], Kun Yao[7], Tom Tiwald[8], Tae Joon Park[6], David P. Landau[5], Haidan Wen[4,9], Subramanian Sankaranarayanan[2,3], Pierre Darancet[2,10], Shriram Ramanathan[6,11] and Yohannes Abate[1*]

[1]Department of Physics and Astronomy, University of Georgia, Athens GA 30602, USA

[2]Center for Nanoscale Materials, Argonne National Laboratory, Lemont, IL 60439, USA

[3]Department of Mechanical and Industrial Engineering, University of Illinois, Chicago, IL, 60607, USA

[4]Advanced Photon Source, Argonne National Laboratory, Lemont, Illinois 60439, USA

[5]Center for Simulational Physics and Department of Physics and Astronomy, University of Georgia, Athens GA 30602, USA

[6]School of Materials Engineering, Purdue University, West Lafayette, IN 47907, USA

[7]School of Electrical and Computer Engineering, University of Georgia, Athens GA 30602, USA

[8]J.A. Woollam Co., Inc., Lincoln, NE 68508, USA

[9]Materials Science Division, Argonne National Laboratory, Lemont, IL 60439, USA

[10]Northwestern Argonne Institute of Science and Engineering, Evanston, IL 60208, USA

[11]Department of Electrical & Computer Engineering Rutgers, The State University of New Jersey, Piscataway, NJ 08854, USA

*Corresponding author E-mail: yohannes.abate@uga.edu

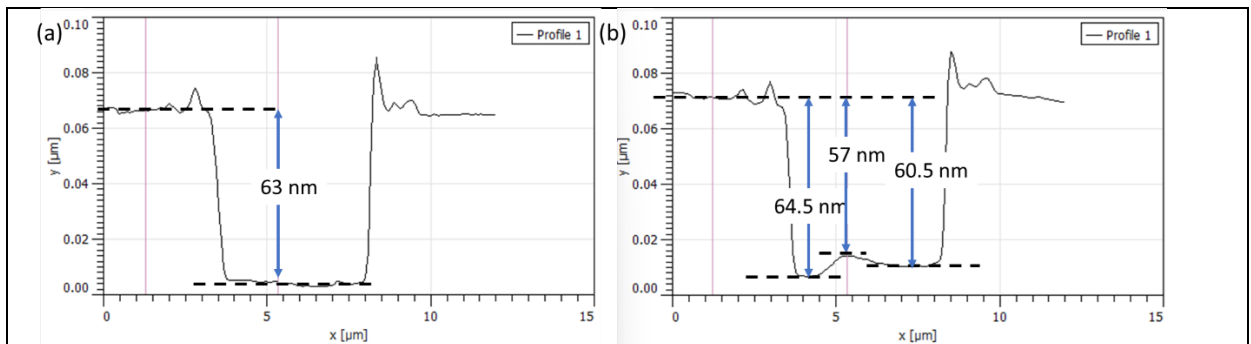

**Figure S1. Expansion of H-NNO film with reference to the electrode under the influence of applied electric field.** (a) Relative height from the H-NNO film to the top of the electrode

without any electric field applied. (b) Relative height from the different regions of the H-NNO film when electric field of 1.46 Vμm⁻¹ is applied from left to right across electrodes.

Increase in thickness (%) of H-NNO film under the E-field was calculated as follows:

    Max. change in height        = (63-57) nm

                               = 6 nm

    Grown thickness of NNO film ≈ 100 nm

    % Increase in thickness        ≈ 6%

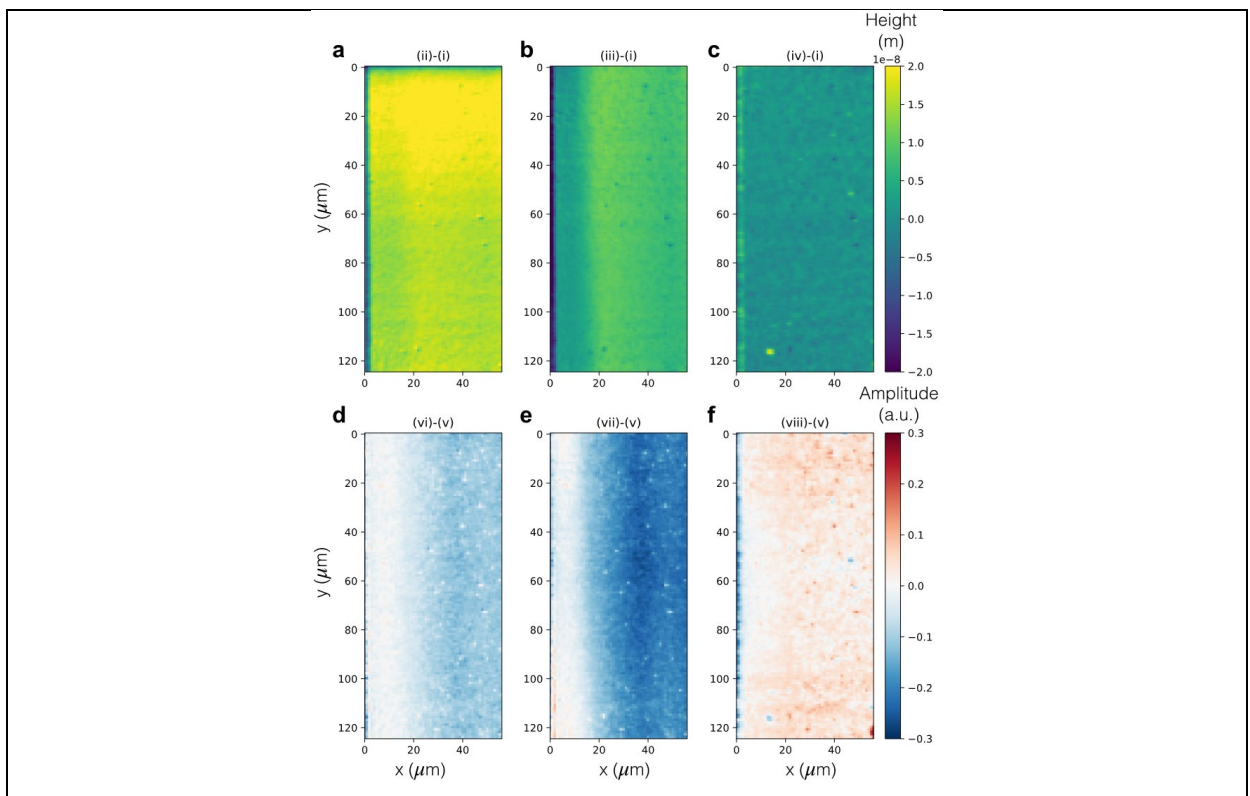

**Fig. S2** Measured changes in height (top) and amplitude (bottom) with respect to scanning position induced by a finite electric field. Top: Difference in the height profiles measured

through topography between E=1.12 Vμm$^{-1}$ (a), 1.46 Vμm$^{-1}$ (b) and final, 0 Vμm$^{-1}$ (c) and the initial, no field configuration. Bottom (d-e): Same for scanning near-field optical microscopy (SNOM) images. Calculated Pearson correlation are 0.44 (between a and d), 0.2 (between b and e),, and 0.56 (between c and f).

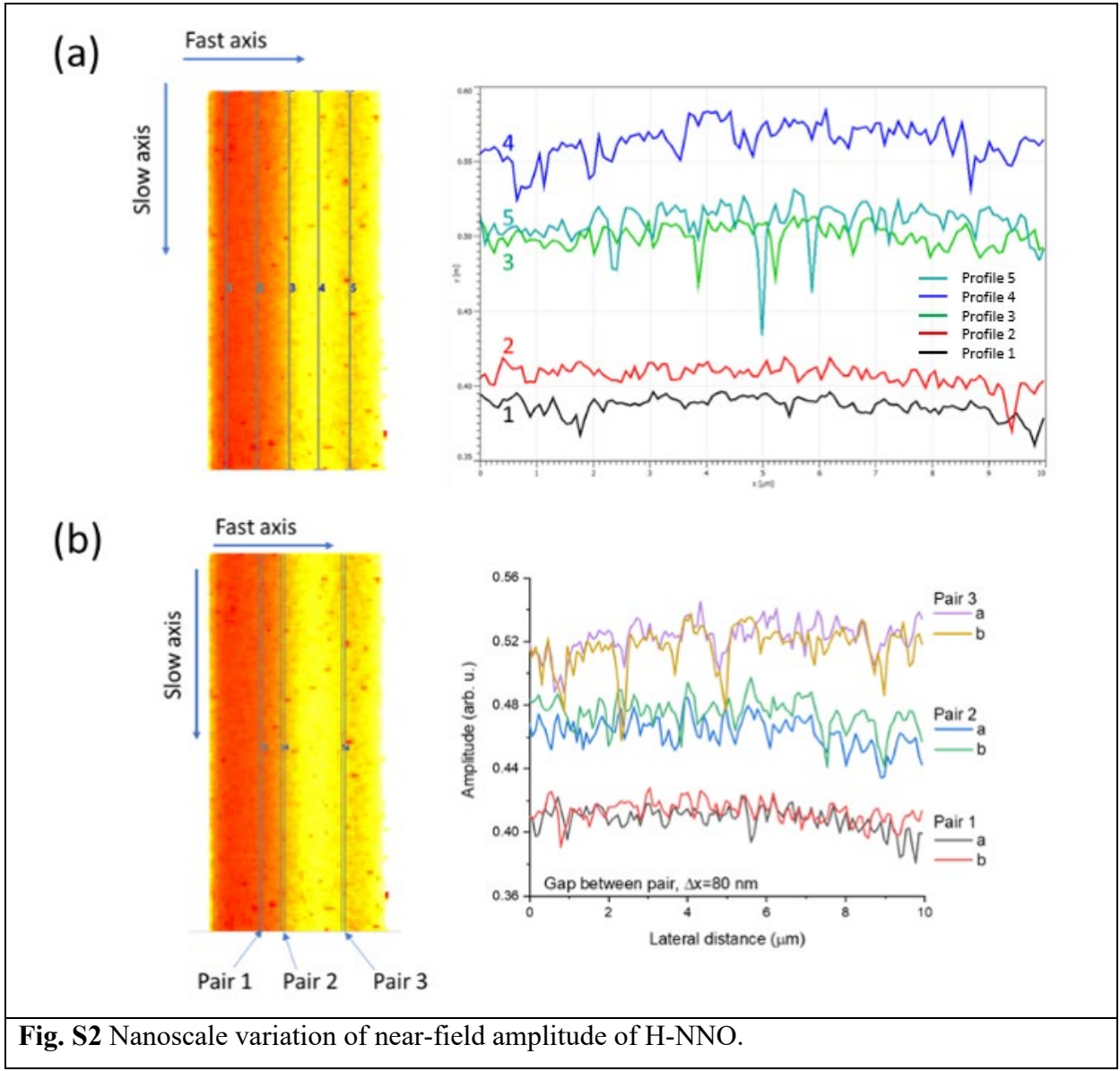

**Fig. S2** Nanoscale variation of near-field amplitude of H-NNO.

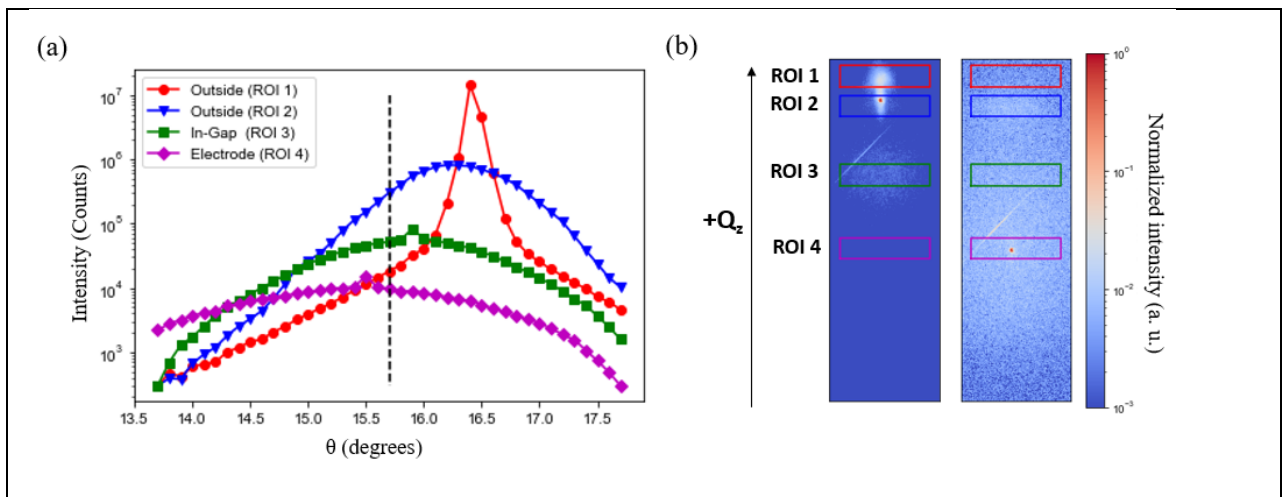

**Fig. S3** HRXRD rocking curves of the Pd-Pd device inside the gap, with a black dashed line indicating the sample angle ($\Theta$ = 15.7°) at which Figure 2 was measured. Rocking curves were taken in 3 different spatial locations: outside the device, in the device gap, and under the device electrode. The intensity at 4 different regions of interest (ROIs) on the detector were summed and plotted as a function of sample angle. (b) Representative x-ray diffraction images taken from the in-gap rocking curves at 16.3° (left) and from rocking curves taken underneath the electrode at 15.5° (right) with the various region of interests (ROIs) that were used in Figure S3a shown as different colored rectangles on the diffraction images. The linear, diagonal feature on the diffraction pattern is diffuse scattering from the substrate.

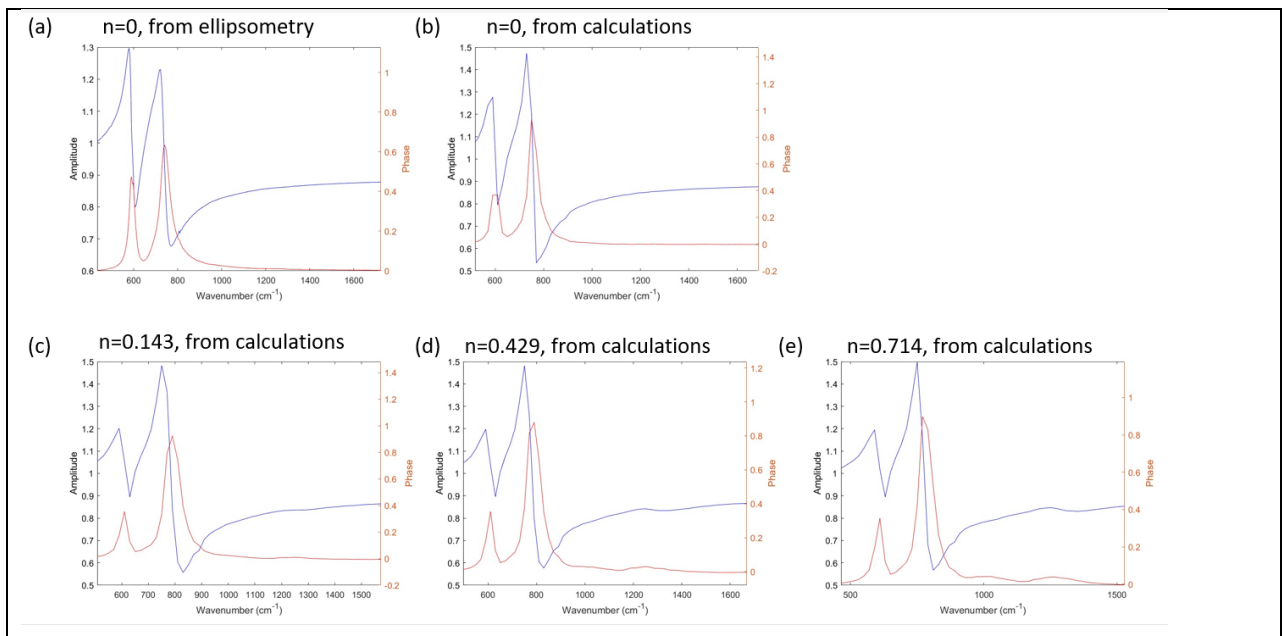

**Fig. S4** Calculated near-field amplitude and phase spectra of NNO with different hydrogen doping levels (*n*), using dipole model.

The near-field amplitude and phase of NNO samples with different doping levels (*n*) were calculated using extended finite dipole model and presented in Figure S4. First, we used the dielectric function values of undoped NNO obtained from ellipsometry to calculate the amplitude and phase and resulted spectra are shown in Fig. S4(a). Then we used the dielectric values obtained from molecular dynamics calculations to generate the amplitude and phase spectra for NNO with doping levels, n=0 (b), 0.143 (c), 0.429 (d) and (e) 0.714.

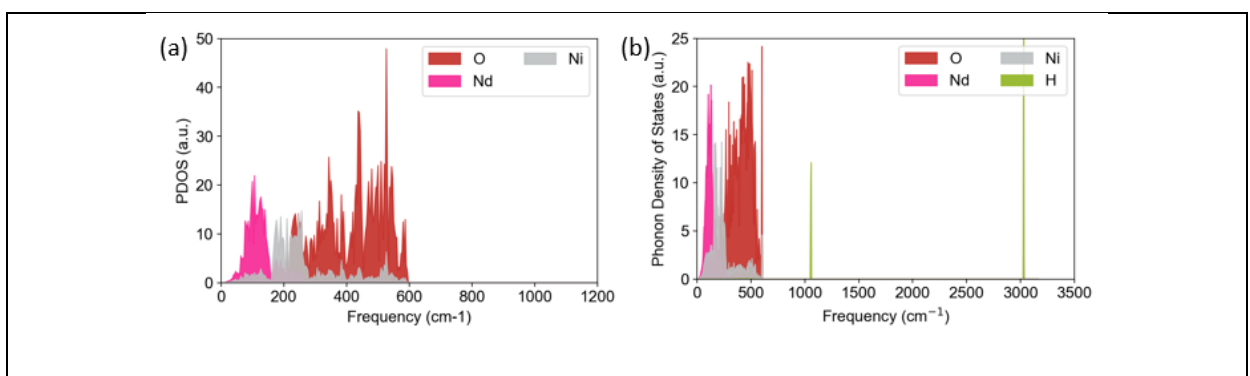

**Fig. S5** (a) Phonon density of states for (a) pristine and (b) $H_{0.25}$-NNO.